\documentclass[prl,twocolumn,aps]{revtex4}
\usepackage{graphicx}
\usepackage{bm}

\newcommand{\be}{\begin{equation}}
\newcommand{\ee}{\end{equation}}
\newcommand{\beq}{\begin{eqnarray}}
\newcommand{\eeq}{\end{eqnarray}}

\begin{document}

\title{Steady-State Chemotactic Response in \textit{E. coli}}

\author{Yariv Kafri$^{1}$ and Rava Azeredo da Silveira$^{2}$}

\affiliation{$^{1}$ Department of Physics, Technion, Haifa 32000, Israel\\
$^{2}$ Department of Physics and Department of Cognitive Studies, \'{E}cole
Normale Sup\'{e}rieure, 24, rue Lhomond, 75005 Paris, France}

\begin{abstract}
The bacterium \textit{E. coli} maneuvers itself to regions with high
chemoattractant concentrations by performing two stereotypical moves:
`runs', in which it moves in near straight lines, and `tumbles', in which it
does not advance but changes direction randomly. The duration of each move
is stochastic and depends upon the chemoattractant concentration experienced
in the recent past. We relate this stochastic behavior to the steady-state
density of a bacterium population, and we derive the latter as a function of
chemoattractant concentration. In contrast to earlier treatments, here we
account for the effects of temporal correlations and variable tumbling
durations. A range of behaviors obtains, that depends subtly upon several
aspects of the system---memory, correlation, and tumbling stochasticity in
particular.
\end{abstract}

\maketitle

Chemotaxis refers to directed motion in response to chemical signals and has
been extensively studied in the bacterium \textit{Escherichia coli} (\textit{E. coli}) \cite{BergBook}. \textit{E. coli} is confined to two stereotypical moves. When its flagellum motors turn counterclockwise (looking at the bacteria from the back), the bacterium moves in near-straight lines termed `runs' whose direction is limited by rotational diffusion. This motion is interrupted by periods of `tumble' which occur when the motors turn clockwise: the bacterium does not translate but instead rotates about itself in a random fashion, and thus reinitializes the direction of the next run. Some amount of correlation between successive run directions yields an average angle shift of $68^{0}$ \cite{BergBook}, as compared with the $90^{0}$ in an uncorrelated case. Tumble durations are short, of the order of $0.1$ s, with respect to runs which last about $1$ s \cite{BergBorwn72,Cell}.

Bacteria modulate their whereabouts in response to their chemical environment. The small size ($\sim 2\mu m$) of \textit{E. coli} rules out
sensing spatial gradients: in the time it takes the bacterium to move by its
own size, chemicals diffuse in a region ten times larger \cite{BergPurcell77}
. Instead, \textit{E. coli} relies upon temporal integration: it calculates
a spatial gradient by integrating the concentration of chemicals over its
recent history; it then uses the resulting quantity to modulate run and
tumble durations. Much work has focused on this `algorithm', namely, on the
filter of temporal integration and on its relation to the probability of run
or tumble. \cite{Block83,Segal86,BergBook}.

With knowledge of this stochastic algorithm, one would like to predict the
distribution of trajectories of a bacterium or, equivalently, the behavior
of a (non-interacting) population. And, in particular, one would like to
elucidate which aspects of the single-bacterium algorithm ensure population
performance. Here, we focus on the steady state and we ask the following
questions. Given a chemoattractant (or chemorepellant) concentration and a
single-bacterium stochastic algorithm, what is the shape of the steady-state
population density? How does it depend upon the details of the
single-bacterium system and which of these are qualitatively relevant?

Because of the single-bacterium stochasticity, the problem may be viewed as
a biased random walk problem. The memory involved in temporal integration
and the variable tumble duration, however, make the problem more difficult
and more interesting. In particular, these induce correlations between run
duration and local bacterium density at the run's starting point, which we
took into account. All these effects yield a rich macroscopic behavior in
the steady state that depends subtly upon the form of the single-bacterium
response filter. In particular, (\textit{i}) the usual bi-lobe filters that
turn temporal integration into spatial comparison may or may not lead to
accumulation in favorable regions, depending upon their shape and the
interplay of time scales \cite{deGennes04,ClarkGrant05}; (\textit{ii})
correlations result in a non-local dependence of the probability density
upon the environment, due to the presence of memory in the dynamics; (\textit{iii}) when tumble is non-instantaneous, bacteria may aggregate in
favorable regions in their tumbling phase. Suprisingly, this last effect
occurs even for filters that are purely local in time. Our results are
derived in one spatial dimension, as in Refs. \cite{deGennes04,ClarkGrant05}, and fodder a long-standing debate \cite{Block83,Segal86,Schnitzer1993,BergBook}; specifically, correlations and
tumble duration variability had been neglected in earlier treatments \cite{Block83,Segal86,Schnitzer1993,BergBook,deGennes04,ClarkGrant05}.

\textit{E. coli} climbs up chemical gradients by modulating run and tumble
durations as a function of chemoattractant concentration, $c$ \cite{BergBook,Cell}. (Henceforth, we use the term `chemoattractant'
indifferently to refer to both chemoattractant and chemorepellent. Below, we
discuss the differences in responses to `positive' and `negative' stimuli.)
Run durations are Poissonian, with probability 
\begin{equation}
\frac{dt}{\tau (t)}=\frac{dt}{\tau _{0}}\left\{ 1-\mathcal{F}\left[ c\right]
\right\}  \label{eq:tumbfreq}
\end{equation}
to switch from run to tumble between times $t$ and $t+dt$ \cite{Cell}. Here, 
$\mathcal{F}\left[ c\right] $ is a functional of the chemical concentration, 
$c(t^{\prime })$, experienced by the bacterium at times $t^{\prime }\leq t$;
it results from a linear temporal filtering followed by a static
rectification non-linearity, as 
\begin{equation}
\mathcal{F}\left[ c\right] =\phi \left( \int_{-\infty }^{t}dt^{\prime
}R(t-t^{\prime })c(t^{\prime })\right) \;,  \label{eq:generalresponseREAL}
\end{equation}
where the functions $\phi \left( \cdot \right) $ and $R(t)$ summarize the
action of the biochemical machinery that processes input signals from the
environment \cite{BergBook}. If $\phi \left( \cdot \right) $ is
non-singular, it may be linearized, as 
\begin{equation}
\mathcal{F}_{\mathrm{lin}}\left[ c\right] =\int_{-\infty }^{t}dt^{\prime
}R(t-t^{\prime })c(t^{\prime }),  \label{eq:linearresponse}
\end{equation}
where an additive constant is absorbed in a redefinition of $\tau _{0}$ (in
Eq. (\ref{eq:tumbfreq})) and a multiplicative constant is absorbed in a
rescaling of $R(t)$. Experimental work \cite{Cell} suggests instead a
thresholding non-linearity \cite{Schnitzer1993}, well fitted by the form 
\begin{equation}
\mathcal{F}_{\mathrm{nlin}}\left[ c\right] =\left[ \mathcal{F}_{\mathrm{lin}} \left[ c\right] \right] _{+},  \label{eq:nonlinearresponse}
\end{equation}
where 
\begin{equation}
\left[ x\right] _{+}=\left\{ 
\begin{array}{cc}
0 & \text{if \ }x\leq 0 \\ 
x & \text{if \ }x>0
\end{array}
\right. .  \label{eq:thresholdlinear}
\end{equation}

The response filter, $R(t-t^{\prime })$, was measured in classic experiments
on wild-type bacteria, in which puffs of chemoattractant were presented to a
single bacterium, effectively replacing $c\left( t^{\prime }\right) $ by a
delta-function which allowed one to resolve for $R\left( t-t^{\prime}\right) 
$ \cite{BergBook, Cell}. These experiments yielded a bimodal shape for $R\left( t-t^{\prime }\right) $, with a positive peak around $t^{\prime}\simeq 0.5\sec $ and a negative peak around $t^{\prime }\simeq 1.5\sec $. The negative lobe is shallower than the positive one and extends up to $t^{\prime }\simeq 4\sec$, beyond which it vanishes and to a good approximation satisfies $\int_{0}^{\infty }R(t^{\prime })dt^{\prime }=0$. The estimated value of $\tau_{0}$ is about $1\sec $ (see Fig. 1 for an  
illustration).

Tumble duration also is modulated stochastically, in close analogy to run
duration behavior \cite{Cell}. Earlier theoretical work has mostly treated
tumble as instantaneous \cite{deGennes04,ClarkGrant05,Schnitzer1993}. We
treat tumble duration as a Poisson variable with rate $1/\tau _{T}$ but, for
the sake of simplicity, we ignore any dependence of the latter upon the
chemical environment. While this comes short of a full description, the mere
allocation of a non-vanishing duration to tumble brings in qualitative
consequences, as discussed below.

The bi-lobe shape of the response filter points to a simple mechanism: it
enables the bacterium to perform a coarse-grained temporal derivative of the
chemical concentration it experiences. If the gradient is positive, then the
run duration tends to increase; if the gradient is negative, then the run
duration tends to decrease (in the linear case of Eq. (\ref{eq:linearresponse})) or is unmodulated (in the threshold-linear case of Eq. (\ref{eq:nonlinearresponse})). However, the connection between simple
arguments such as this and quantitative results is far from immediate.
Reference \cite{deGennes04} argues that a single-lobe, even punctual
temporal filter, as $R(t-t^{\prime })=\chi \delta (t-t^{\prime })$ with $\chi $ \textit{positive}, leads to a net bias toward increasing chemoattractant concentration. In fact, the analysis suggests that the response is strongest if the filter is local in time, with $t^{\prime }=0$, and that a \textit{delayed} response ($t^{\prime }>0$) or any addition of a \textit{negative} contribution, akin to the bi-lobe shape measured experimentally, \textit{weakens} the bias. The arguments developed in Ref. \cite{deGennes04} concern the instantaneous dynamics of a bacterium and
ignore the spatially varying buildup of probability in time; they apply, for
example, to a transient situation in which the probability density is flat.
Reference \cite{ClarkGrant05} contrasts transient and steady-state behavior
and argues that while a positive filter is most favorable for climbing
chemical gradients in an initial transient phase, a negative filter is
favorable for steady-state accumulation in advantageous regions. Finally, it
argues that the typical bi-lobe shape of \ the linear filter, $R(t)$, may
derive from a constrained optimization involving transience and steady
state. While Refs. \cite{deGennes04,ClarkGrant05} present a number of
interesting ideas and go some length into explaining chemotaxis
statistically, they make a number of limiting assumptions. First, they
disregard the correlation between run duration and probability density in a
given region. Second, they assume instantaneous tumble. Third, the
constrained optimization in Ref. \cite{ClarkGrant05} is somewhat \textit{ad hoc}.

In the remainder of the paper, we proceed as follows. First, we write
equations that govern the steady-state density of (non-interacting) bacteria
(equivalently, the bacterium probability density); second, we derive the
latter analytically in the linear model (Eq. (\ref{eq:linearresponse})) and
numerically in the non-linear model (Eq. (\ref{eq:nonlinearresponse})). We
are after the density 
\begin{equation}
N(x)=N_{R}(x)+N_{T}(x)\;,
\end{equation}
where $N_{i}(x)dx$ is the number of bacteria lying between $x$ and $x+dx$ in
the steady-state, and the subscripts $R$ and $T$ refer to run and tumble
respectively. As a natural way to incorporate correlation, we borrow four
intermediate quantities: $n_{+}^{T\rightarrow R}(x)dx$, the number of
bacteria that switch from tumble to \textit{rightward} run between $x$ and $x+dx$ per unit time; $n_{-}^{T\rightarrow R}(x)dx$, the number of bacteria
that switch from tumble to \textit{leftward} run between $x$ and $x+dx$ per
unit time; $n_{+}^{R\rightarrow T}(x)dx$, the number of bacteria that switch
from \textit{rightward} run to tumble between $x$ and $x+dx$ per unit time; $n_{-}^{R\rightarrow T}(x)dx$, the number of bacteria that switch from \textit{leftward} run to tumble between $x$ and $x+dx$ per unit time. The rightward and leftward fluxes are given by 
\begin{eqnarray}
\partial _{x}j_{+}(x) &=&n_{+}^{T\rightarrow R}(x)-n_{+}^{R\rightarrow T}(x),
\label{flux-plus} \\
\partial _{x}j_{-}(x) &=&n_{-}^{T\rightarrow R}(x)-n_{-}^{R\rightarrow T}(x)
\text{ }.
%\cite{steadystate-constraint} 
\label{flux-minus}
\end{eqnarray}
If $N_{+}(x)$ and $N_{-}(x)$ are the densities of rightward and leftward
running bacteria respectively, then $j_{+}(x)=vN_{+}(x)$ and $j_{-}(x)=-vN_{-}(x)$, and the total density of running bacteria, $N_{R}(x)$,
obeys 
\begin{eqnarray}
\partial _{x}N_{R}(x) &=&\partial _{x}\left[ N_{+}(x)+N_{-}(x)\right]  \\
&=&\frac{1}{v}\left[ n_{+}^{T\rightarrow R}(x)-n_{+}^{R\rightarrow T}(x)\right.   \nonumber \\
&&\left. -n_{-}^{T\rightarrow R}(x)+n_{-}^{R\rightarrow T}(x)\right] \;.
\label{running-density}
\end{eqnarray}
Tumbling bacteria retain some memory of their recent run direction; we call $q$ the probability that a tumble causes a run direction change, and treat it
as a parameter in our model. Thus, $n_{+}^{T\rightarrow R}(x)=\left(1-q\right) n_{+}^{R\rightarrow T}(x)+qn_{-}^{R\rightarrow T}(x)$ and $n_{-}^{T\rightarrow R}(x)=qn_{+}^{R\rightarrow T}(x)+\left( 1-q\right)n_{-}^{R\rightarrow T}(x)$, so that Eq. (\ref{running-density}) simplifies into 
\begin{equation}
\partial _{x}N_{R}(x)=\frac{2q}{v}\left[ n_{-}^{R\rightarrow
T}(x)-n_{+}^{R\rightarrow T}(x)\right] .
\end{equation}
Within our assumption of unmodulated tumble rate, in the steady state the
density of tumbling bacteria, $N_{T}(x)$, reads 
\begin{equation}
N_{T}(x)=\tau _{T}\left[ n_{+}^{R\rightarrow T}(x)+n_{-}^{R\rightarrow T}(x)
\right] .
\end{equation}

As a final simplifying assumption, we posit that memory is erased at
tumble-to-run switches. This assumption may not be validated by data \cite
{Cell}, but it is unclear whether it improves or suppresses chemotaxis with
respect the no-erasure case. Equation (\ref{eq:linearresponse}) becomes 
\begin{equation}
\mathcal{F}_{\mathrm{lin}}\left[ c\right] =\int_{t_{0}}^{t}dt^{\prime
}R(t-t^{\prime })c(t^{\prime }),
\end{equation}
where $t_{0}$ is the time of last switch, and the run-to-tumble switch
probability, $dt/\tau (t,t_{0})$, is now a function of both $t$ and $t_{0}$.
Alternatively, this probability can be expressed in terms of the initial and
final positions of the run, $y$ and $x$ respectively, as $dx/v\tau (x,y)$.
We now have all the elements in hand to write the steady-state equations
that govern density and keep track of correlations, as 
\begin{eqnarray}
n_{+}^{R\rightarrow T}(x) &=&\int_{-\infty }^{x}dyn_{+}^{T\rightarrow
R}(y)\rho _{+}(x,y),  \label{eq:maineq1} \\
n_{-}^{R\rightarrow T}(x) &=&\int_{x}^{+\infty }dyn_{-}^{T\rightarrow
R}(y)\rho _{-}(x,y);  \label{eq:maineq2}
\end{eqnarray}
these express the fact that tumbling bacteria result from running bacteria
that switch to tumbling mode. Here, $\rho _{+}(x,y)dx$ and $\rho _{-}(x,y)dx$
are probabilities that a bacterium, which tumbled last at $y$, tumbles again
between $x$ and $x+dx$ (and not before), for $x>y$ and $x<y$ respectively.
These probabilities are given by 
\begin{equation}
\rho _{\pm }(x,y)dx=\exp \left( \mp \int_{y}^{x}dy^{\prime }\frac{1}{v\tau
(y^{\prime },y)}\right) \frac{dx}{v\tau (x,y)}.  \label{eq:rhot}
\end{equation}

We choose to illustrate our results with steps of chemoattractant
concentration, $c(x)=\xi \theta (x)$ ($\xi >0$), where $\theta (x)$ denotes
the Heaviside function. In the linear model, it is handy to focus upon
singular response functions with $R_{\Delta }(t)\propto \delta (t-\Delta /v)$
, or equivalently in space coordinates, $R_{\Delta }(x)=\chi _{\Delta}\delta
(x\mp \Delta )$ (with a minus (plus) sign for rightward (leftward) runs).
One can then derive solutions for more general cases as linear
superpositions of solution for singular response functions. We treat the
linear model perturbatively in the strength of bacterium response;
specifically, we assume a regime with $\alpha _{\Delta }\equiv \xi
\chi_{\Delta }\ll 1$. (For $\alpha _{\Delta }=0$, there is no chemotaxis.)
Expanding Eq. (\ref{eq:rhot})\ to first order in $\alpha _{\Delta }$, we
solve the steady-state Eqs. (\ref{eq:maineq1}) and (\ref{eq:maineq2}) for
the intermediate quantities $n_{\pm }^{R\leftrightarrow T}$. From these, we
derive the incremental running and tumbling bacterium densities compared to
the densities far to the left of the chemoattractant step: $\delta
N_{R}^{\Delta }(x)\equiv N_{R}^{\Delta }(x)-N_{R}^{\Delta }(-\infty )$ and $
\delta N_{T}^{\Delta }(x)\equiv N_{T}^{\Delta }(x)-N_{T}^{\Delta }(-\infty )$.

Because of the singular response function and the discontinuity in
chemoattractant density at $x=0\,$, our solutions have singular points at $
x=\pm \Delta $. We find, for $x<-\Delta $, 
\begin{equation}
\delta N_{R}^{\Delta }(x)=\delta N_{T}^{\Delta }(x)=0;
\end{equation}
for $-\Delta \leq x\leq \Delta $, 
\begin{eqnarray}
\delta N_{R}^{\Delta }(x) &=&-2aq\frac{\alpha _{\Delta }(x+\Delta)}{v^{2}\tau _{0}}
e^{-\Delta /v\tau _{0}},  \label{eq:sol1-R} \\
\delta N_{T}^{\Delta }(x) &=&-a\frac{\alpha _{\Delta }\tau _{T}}{v\tau _{0}}
\left( 1+2q\frac{(\Delta +x)}{v\tau _{0}}\right) e^{-\Delta /v\tau _{0}};
\label{eq:sol1-T}
\end{eqnarray}
for $x>\Delta ,$ 
\begin{eqnarray}
\delta N_{R}^{\Delta }(x) &=&-4aq\frac{\alpha _{\Delta }\Delta }{v^{2}\tau
_{0}}e^{-\Delta /v\tau _{0}},  \label{eq:sol2-R} \\
\delta N_{T}^{\Delta }(x) &=&-a\frac{\alpha _{\Delta }\tau _{T}}{v\tau _{0}}
\left( 1+4q\frac{\Delta }{v\tau _{0}}\right) e^{-\Delta /v\tau _{0}} 
\nonumber \\
&=&\left( 2+\frac{v\tau _{0}}{2q\Delta }\right) \frac{\tau _{T}}{\tau _{0}}
\delta N_{R}^{\Delta }\left( x\right) ;  \label{eq:sol2-T}
\end{eqnarray}
here $a$ is positive a constant that sets the overall density of bacteria.
From Eq. (\ref{eq:sol2-R}), running bacteria accumulate to the right if $
\alpha _{\Delta }<0\,$, as long as the `response memory' is non-vanishing ($
\Delta \neq 0$). Accumulation is strongest for $\Delta =v\tau _{0}$, \textit{
i. e.} , when the response memory, $\Delta /v$, is comparable to the typical
run duration, $\tau _{0}$. We note also that accumulation vanishes if $q=0$;
indeed, in this case bacteria do not change their run direction after tumble
and, hence, behave roughly as if there were no tumbles whatsoever. As
typically $\tau _{T}\ll \tau _{0}$ (experimentally, for \textit{E. coli}, $
\tau _{T}\approx \tau _{0}/10$), Eq. (\ref{eq:sol2-T}) implies that $\delta
N_{T}^{\Delta }$ is dominated by $\delta N_{R}^{\Delta }$. However, the
reverse occurs in the particular case with small response memory $\Delta
/v<\tau _{T}/2$, \textit{i. e.}, when the typical tumble duration exceeds
the response memory. In this case, bacteria may accumulate to the right (if $
\alpha _{\Delta }<0$) \textit{even for a response function purely local in
time} (with $\Delta =0$)---a possibility overlooked in earlier work that
treat tumble as instantaneous. In this tumbling-dominated regime, bacteria
accumulate at favorable tumbling sites while the uniformly populated runs
serve as a way to explore potentially favorable tumbling positions.

\begin{figure}[th]
\center
\includegraphics[width=6cm]{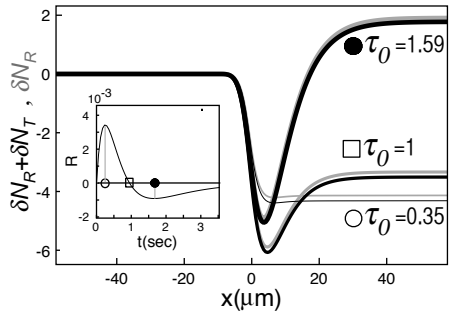}
\caption{Numerical results in the non-linear model with $c(x)=10^{-3} 
\protect\theta(x)$. All quantities are given in arbitrary units. To obtain $
\protect\delta N_R(x)$ and $\protect\delta N_T(x)$, Eqs. (\protect\ref
{eq:maineq1}-\protect\ref{eq:maineq2}), for the non-linear model, were
solved iteratively on a computer, using a discrete lattice of size 800, with 
$v=10 \protect\mu m$, $q=0.4$, $\protect\tau_T=10/17 s$ and $dx=0.05 \protect
\mu m$. The results were uniformly rescaled for convenience and we display
the region in which the bacterium density varies, about $x=0$, the location
of the chemoattractant step. Three different values of $\protect\tau_0$ (in
seconds) are indicated in the figure. The correspondence between the
bacterium density and the value of $\protect\tau_0$ is indicated by the
solid and open symbols, in both figure and inset in which the response
function is illustrated. The functional form of the response function was
chosen as $R(t)=(240 t \exp(-200t/17)-29.4 t \exp(-7t/17))/17$ which
satisfies $\protect\int_0^\infty R(t) dt=0$.}
\label{fig}
\end{figure}

Our analysis suggests that bacteria accumulate in favorable regions if the
impulse response function is negative. As remarked in Ref. \cite
{ClarkGrant05}, this conclusion is paradoxical in view of experimental
measurements, which yield a bi-lobe response function \cite{Cell}. For
comparable positive and negative lobes, chemotaxis ought to work best if the
negative lobe is peaked around a time $\tau _{0}$ in the past, and fail if
it is relegated much beyond in the past. We illustrate this issue in Fig. 1,
where we plot solutions of the \textit{non-linear} model (Eq. (\ref
{eq:nonlinearresponse})) for a step of chemoattractant concentration. We use
a bi-lobe response function similar to the experimental one and derive the
steady-state density of bacteria for three different values of $\tau _{0}$.
According to Fig. 1, accumulation in favorable regions occurs when $\tau
_{0} $ is comparable to the time of the negative peak in the response
function (top curve in Fig. 1 labeled by a disk symbol). For smaller values
of $\tau _{0}$, bacteria feel the negative peak only rarely and accumulation
occurs in unfavorable regions. This picture agrees with our analytical
results in the linear model.

Curiously, the experimental value $\tau _{0}$ generally quoted ($\sim $1 $
\sec $) falls between the two peaks of the response function and, in our
model, does not lead to favorable accumulation (intermediate curve in Fig. 1
labeled by an open square). This conclusion may be modified for a different
shape of the response function, less similar to the experimental one---for
example, one with a very deep negative lobe. Obviously, there are a number
of constraints and performance requirements which we have not considered and
which inform the shape of the single-bacterium filter. For example, a
rational for a response function that is spread out in time instead of
narrowly peaked is the resulting robustness with respect to input noise, and
a rational for a bi-lobe response function is the resulting `adaptive'
mechanism of mean subtraction. 

In sum, we have introduced steady-state equations that govern bacterium
density in chemotactic response to a chemoattractant profile. The solutions
present a rich behavior which depends in a subtle manner on the details of
the model. We find that the bacterium density is a non-local function of the
chemoattractant density (see Eqs. (\ref{eq:sol1-R}--\ref{eq:sol2-T}) and
Fig. 1). This feature of the steady state is a direct consequence of the
presence of memory in the dynamics and emerges in a proper treatment of
correlations; earlier studies which ignore correlations find local solutions 
\cite{ClarkGrant05}. Our approach also predicts a regime in which bacteria
accumulate favorably, even in the case of memory-less dynamics, in the
tumbling state. Most earlier studies treat tumble as instantaneous. We
treated tumble duration as a homogeneous Poisson process. In experiments,
tumble duration seems to be influenced by the recent past in much the same
way as run duration is, but with a bi-lobe response function that is more
narrowly peaked and sign-inverted \cite{Cell}. Roughly, we may say that
tumbles tend to be shorter in favorable regions and longer in unfavorable
regions. If so, chemotactic response may be weakened by this effect, with
respect to the homogeneous tumble case.

Acknowledgments. We thank H. Berg, D. A. Clark, L. C. Grant, D. Levine, and
A. Samuel for useful discussions. I. Kirjner is thanked for valuable
contributions on the tumbling-dominated regime. YK is supported by the
Israel Science Foundation. RAS is supported in part by the CNRS through UMR
8550.


\begin{thebibliography}{99}
\bibitem{BergBook} H. C. Berg, \textit{E. Coli in Motion} (Springer, New
York 2004).

\bibitem{BergBorwn72} H. C. Berg and D. A. Brown, Nature, \textbf{239}, 500
(1972).

\bibitem{Cell} S. M. Block, J. E. Segall and H. C. Berg, Cell, \textbf{31},
215 (1982).

\bibitem{BergPurcell77} H. C. Berg and E. M. Purcell, Biophys. J., \textbf{\
20 }, 193 (1977).

\bibitem{Block83} S. M. Block, J. E. Segall and H. C. Berg, J. Bacteriol. 
\textbf{154}, 312 (1983).

\bibitem{Segal86} J. Segall, S. Block and H. C. Berg, Proc. Natl. Acad. Sci.
(USA), \textbf{83}, 8987 (1986).

\bibitem{deGennes04} P.-G. de Gennes, Eur. Biophys. J, \textbf{33}, 691
(2004).

\bibitem{ClarkGrant05} D. A. Clark and L. C. Grant, Proc. Natl. Acad. Sci.
(USA), \textbf{102}, 9150 (2005).

\bibitem{Schnitzer1993} M. J. Schnitzer, Phys. Rev. E, \textbf{48}, 2553
(1993).

%\bibitem{steadystate-constraint} In the steady state, the absence of
%accumulation or depletion of tumbling bacteria implies $n_{+}^{T\rightarrow
%R}(x)+n_{-}^{T\rightarrow R}(x)=n_{+}^{R\rightarrow
%T}(x)+n_{-}^{R\rightarrow T}(x)$. From Eqs. (\ref{flux-plus}, %\ref{flux-minus}), this identity coincides, as expected, with the usual
%steady-state condition, $\partial _{x}\left[ j_{+}(x)+j_{-}(x)\right] =0$.

\bibitem{tobepublished} Y. Kafri and R. Azeredo da Silveira, in preparation.

\end{thebibliography}
\end{document}